# Locating Crashing Faults based on Crash Stack Traces

Liang Gong[†], Hongyu Zhang[†], Hyunmin Seo[§] and Sunghun Kim[§]

[†]School of Software, Tsinghua University
Beijing 100084, China
gongliang10@mails.tsinghua.edu.cn, hongyu@tsinghua.edu.cn

[§]Department of Computer Science and Engineering
The Hong Kong University of Science and Technology, Hong Kong, China
hunkim@cse.ust.hk

*Abstract*—Software crashes due to its increasing complexity. Once a crash happens, a crash report could be sent to software developers for investigation upon user permission. Because of the large number of crash reports and limited information, debugging for crashes is often a tedious and labor-intensive task. In this paper, we propose a statistical fault localization framework to help developers locate functions that contain crashing faults. We generate the execution traces for the failing traces based on the crash stack, and the passing traces from normal executions. We form program spectra by combining generated passing and failing trace, and then apply statistical fault localization techniques such as *Ochiai* to locate the crashing faults. We also propose two heuristics to improve the fault localization performance. We evaluate our approach using the real-world Firefox crash report data. The results show that the performance of our method is promising. Our approach permits developers to locate 63.9% crashing faults by examining only 5% Firefox 3.6 functions in the spectra.

## I. INTRODUCTION

Software crash is one of the most severe manifestations of software faults. Crashes need to be fixed with a high priority. Unlike other faults, relevant information about crashing faults can be automatically collected and reported at the time of crash. Many crash reporting systems such as Windows error reporting [9], Apple crash report [2], and Mozilla crash report [23] have been proposed and deployed.

The first step of debugging crashing faults is to locate the faults, but it is still remained as manual tasks. Most existing crash reporting systems [9] [2] [23] are focusing on colleting crash reports, bucketing (classifying) and prioritizing them. Although the collected crash report information is useful for debugging, these systems do not support automatic crashing fault localization.

On the other hand, many statistical fault localization techniques [13] [26] [17] [18] [5] have been proposed to guide developers locate faults. Existing statistical fault localization methods suggest a list of suspicious program entities using program spectra [13], which consist of both passing and failing execution traces. The developers can then examine the ranked list and locate the faults.

Unfortunately, conventional statistical fault localization techniques become impractical for locating crashing faults, because crash reports typically only contain call stack trace [4] rather than the complete failing and passing execution traces. Recovering complete failing execution traces from crash reports is tedious, time-consuming and sometimes infeasible, especially when the reported crash is difficult to reproduce [3]. In addition, crash reports do not contain any passing execution traces.

In this paper, we propose a new framework, which aims to locate post-release crashing faults based on call stack traces. Our approach expands crash stack traces into failing execution traces according to static function call graph extracted from source code. Our approach also generates passing execution trace by executing test cases. Then statistical fault localization techniques such as *Ochiai* are applied on the synthesized execution traces to determine suspicious functions and guide debugging. To improve the fault localization performance, two heuristics are proposed: distance reweighting and test coverage reweighting.

We evaluate our framework using actual crash data from the Firefox 3.6 and 4.0 projects. The evaluation results are promising: for Firefox 3.6, our approach can locate more than 63.9% of crashing faults by examining 5% of functions. For Firefox 4.0, our approach can locate more than 52.7% of crashing faults by examining 5% of functions.

The main contribution of this paper is in twofold:

- We propose a novel framework for locating crashing faults. Our framework is based on crash track traces only and does not require any additional information from the field. To our best knowledge, this is the first time such a framework is proposed.
- We evaluate our approach on Firefox, which is a real world and large-scale project.

The remainder of this paper is organized as follows: We introduce the background information in Section II. Section III describes the challenges for post-release crashing fault localization. Section IV describes our approach to fault localization based on crash stack traces. Section V presents our experimental design, and Section VI shows the experimental results. We discuss issues involved in our approach in Section VII and threats to validity in Section VIII. Section IX surveys related work followed by section X that concludes this paper.

## II. BACKGROUNDS

### A. Crash Reporting System

Despite tremendous efforts on software quality assurance, released software products are often shipped with faults. Some faults manifest as crashes in the field. The crash information from the field can be very useful to fix the defects quickly.

To collect crash information, many Crash-Reporting Systems (CRS) such as Windows error reporting [9], Apple crash report [2], and Mozilla crash report [23] have been proposed and widely deployed. When a crash happens in the field, the CRS collects crash information such as product name, version, time, crashed method signature and call stack trace. Then, the crash information is sent to the server side upon user permission. The server checks the duplication of crash report and assigns it to a bucket, which is typically a set of crashes with the same method signature (the top frame in the call stack trace) [9]. Finally, the CRS presents the crash report to developers.

Figure 1 shows an example of crash stack trace in a Firefox crash report (crash ID: *c5e5bcaf-99fb-48ab-805b-4154c2110315*). The program was crashed at *Frame 0*. Each frame contains a full-qualified function name and its parameters. In real-world, large-scale systems such as Microsoft Windows and Mozilla-Firefox, developers receive a large number of crash reports sent from users all over the world. For example, Mozilla receives on peak day 2.5 million crash reports[1].

Given the large number of crash reports, it is often time-consuming and tedious for developers to locate the root cause of all crashes manually. To facilitate debugging, we propose a statistical fault localization approach based on the call stack trace collected by a CRS.

| Frame 0 (Crash Signature) | nsXULDocument::OnStreamComplete(nsIStreamLoader*,nsISupports*, unsigned int,unsigned int,unsigned char const*) |
|---|---|
| Frame 1 | nsStreamLoader::OnStopRequest(nsIRequest*,nsISupports*,unsigned int) |
| Frame 2 | nsJARChannel::OnStopRequest(nsIRequest*,nsISupports*,unsigned int) |
| Frame 3 | nsInputStreamPump::OnStateStop() |
| ... | |
| Frame 11 | __tmainCRTStartup |
| Frame 12 | BaseProcessStart |

Figure 1. A crash stack example.

### B. Statistical Fault Localization – A General Process

Statistical fault localization techniques facilitate debugging by allowing programmers to examine a small portion of code for locating faults. These techniques usually contrast the program spectra information [20] such as execution traces between passing and failing executions, and compute the fault suspiciousness of individual program elements (such as statements and methods). Finally, they present a list of program elements ranked by their fault suspiciousness. Developers can then identify faults by examining the ranked program elements. Figure 2 shows a general process of a typical statistical fault localization method. Some major steps are described as follows:

**Program instrumentation**: For statistical fault localization, the first step is collecting program spectra by instrumenting the target programs. The instrumentation granularity (e.g. predicate level [17], statement level [10] [13], branch level [6] etc.) determines the granularity of fault localization. After program instrumentation, code that records the execution traces is compiled into target programs.

**Collecting execution traces**: This step executes the instrumented program with test suites and collects the passing and failing execution traces as program spectra.

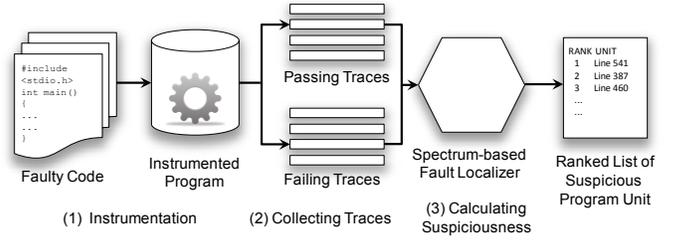

Figure 2. A general statistical fault localization process.

**Calculating suspiciousness**: Collected passing and failing execution traces are used to compute fault suspiciousness scores. Higher suspiciousness score indicates that the corresponding program entity is more likely to contain a fault. Finally, the program elements are ranked by the suspiciousness scores in the descending order, and the ranked elements are presented to developers to guide their debugging process.

Over the years, many statistical fault-localization methods [15] have been proposed. *Tarantula* [12] [13] is one of the first statistical fault localization techniques. It proposed formulas to rank statements in C programs based on the percentages of failing and passing test cases that execute the statements. *Jaccard* [1] and *Ochiai* [1] generally follow the process described in Figure 2, but differ in the way they compute suspiciousness scores. Empirical studies [1] [13] [19] [33] show that these techniques are effective in guiding programmers to examine code and locate faults. We list these statistical fault localization techniques in Table I.

TABLE I. CONVENTIONAL STATISTICAL FAULT LOCALIZATION TECHNIQUES

| Name | Formula |
|---|---|
| Ochiai | $\dfrac{a_{ef}}{\sqrt{(a_{ef}+a_{nf})\cdot(a_{ef}+a_{ep})}}$ |
| Tarantula | $\dfrac{\frac{a_{ef}}{a_{ef}+a_{nf}}}{\frac{a_{ef}}{a_{ef}+a_{nf}}+\frac{a_{ep}}{a_{ep}+a_{np}}}$ |
| Jaccard | $\dfrac{a_{ef}}{a_{ef}+a_{nf}+a_{ep}}$ |

For each function $f_i$, the formulas in Table I calculates the suspiciousness score of $f_i$ containing faults. $a_{ef}$ denotes the number of failing traces that executed function $f_i$. Similarly, $a_{ep}$ is the number of passing traces that executed $f_i$. $a_{np}$ and $a_{nf}$ represents the number of passing and failing traces that didn't execute the function. Basically, if elements are covered more in the failing traces, but not in the passing traces, the elements will get higher suspiciousness scores.

### III. CHALLENGES FOR LOCATING CRASHING FAULTS

Existing fault localization techniques described in Section II.B require complete passing and failing execution traces. These techniques cannot be directly applied to locate crashing faults based on crash stack traces, for two main reasons: 1) they only have partial crash (failing) traces, and 2) they do not contain passing program execution traces.

---
[1] http://blog.mozilla.com/webdev/2010/05/19/socorro-mozilla-crash-reports/

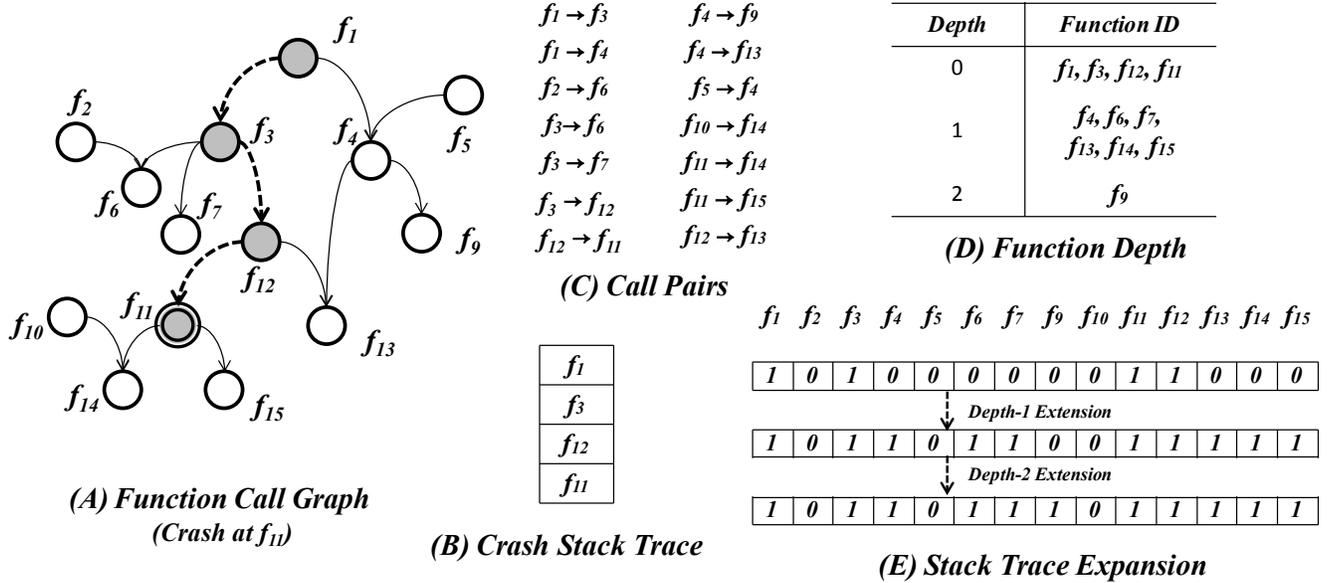

Figure 3. Stack trace expansions.

For example, Figure 3(A) shows a function call graph of a simple program, each node represents a function ($f_i$). Suppose the program starts from $f_1$ and crashes at $f_{11}$. The crash call stack trace (Figure 3(B)), $f_1 \to f_3 \to f_{12} \to f_{11}$ is incomplete, while a complete failing trace could be $f_1 \to f_3 \to f_6 \to f_{12} \to f_{11} \to f_{14}$ and a passing trace could be $f_1 \to f_3 \to f_6 \to f_{12} \to f_{13}$.

To obtain complete execution traces, it is possible to deploy an instrumented version to the end users. The instrumented version could monitor program execution in field and send the passing and complete traces to developers. However, this monitoring often slows down subject systems significantly. For example, [22] [29] analyzed dynamic binary instrumentation overhead, and found that the average performance overhead is around 30%-150% for simple instrumentation, which is unacceptable to end users.

Alternatively, developers may reproduce crashes in house and capture the complete traces when they reproduce it to recover a complete failing execution trace from a call stack trace. However, reproducing crashes often requires non-trivial effort [3].

In summary, conventional fault localization techniques require passing and failing traces. However, crash stack traces do not include complete failing execution traces and passing traces. Therefore, it is impractical to apply conventional fault localization techniques to crashes that are reported by crash reporting systems. We describe our approach to address these challenges in the next section.

## IV. PROPOSED APPROACH

### A. Overall Framework

In this section, we propose a novel statistical fault localization framework for locating crashing faults based on crash stack traces. Our approach addresses the challenges described in Section III by inferring approximated failing execution traces from stack traces. To collect passing traces, we execute instrumented software in house. In addition, we propose two heuristics to improve fault localization performance. The overall process is shown in Figure 4.

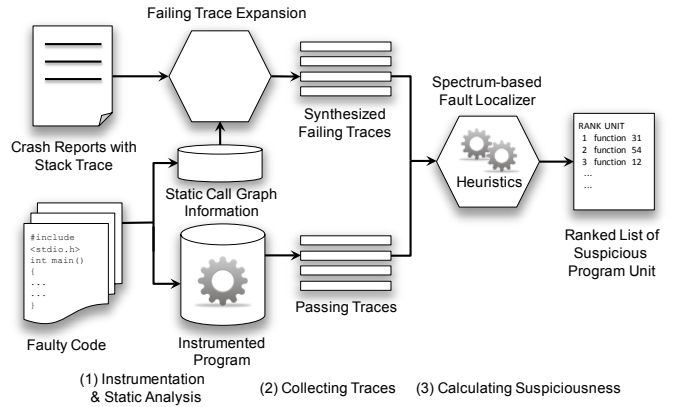

Figure 4. Proposed fault localization framework based on crash stack trace.

### B. Generation of passing and failing execution traces

In our approach, we generate passing execution traces by instrumenting programs and executing them using test suits. We use an instrumentation framework, Pin [22], to intercept the program execution by inserting logging code at the entry point of each function.

Instrumented programs are executed with test suits provided by the project as input. We collect execution traces from all non-crashing test suits, and assume that they are the passing execution traces.

To generate complete failing traces, we first construct a static function call graph [25], which captures all possible call relationship between functions. Each *call pair* consists of the caller and callee functions. As an example, Figure 3(A) shows a static function call graph, and Figure 3(C) shows 14 call pairs that can be obtained by program analysis.

The call graph is used in the algorithm, EXPAND-STACK-TRACE, to generate approximate failing execution traces from stack trace. Before explaining the algorithm, we define a concept called *Call Depth*:

**Definition:** the *Call Depth* of function $f_i$ is the least number of function call steps from any functions in the stack trace.

As an example, Figure 3(D) shows the Call Depths of all functions in Figure 3(A). $f_1$ is in the stack trace so its depth is 0. $f_{13}$ can be called either by $f_4$ in 2 steps ($f_1 \rightarrow f_4 \rightarrow f_{13}$) or by $f_{12}$ in 1 step ($f_{12} \rightarrow f_{13}$). In this case, its Call Depth is 1 as we choose the least number of steps as the call depth. Figure 3(D) lists all function depths in Figure 3(A).

EXPAND-STACK-TRACE(S,d)
```
1: create HashSets Set_0, Set_1, Set_2, ..., Set_d
2: for each function f_i ∈ S do
3:     insert f_i into Set_0
4: end for
5: for k ← 1 to d do
6:     for each function f_j ∈ Set_k do
7:         P ← all call pairs starts from f_j
8:         for each pair <f_j, f_x> ∈ P do
9:             add f_x into Set_{k+1}
10:        end for
11:    end for
12: end for
13: Set_all ← ⋃_{k←0}^{d} Set_k
14: mark all functions in Set_all hit in execution trace t
15: return t
```

The EXPAND-STACK-TRACE algorithm iteratively expands the stack as shown above. It takes $S$ and $d$ as inputs, where $S$ is a crash stack trace, and $d$ is a predefined number of depths to expand. First, we mark all functions in the original crash stack trace as depth-0. Then for each function $f$ in depth-0, all functions that are called by $f$ are added into depth-1. This process is repeated in the **for** loop at line 5 until $k$ reaches the predefined depth $d$. Finally, all functions from all depths are collected to obtain the final set of functions that are hit by the execution trace.

### C. Locating Crashing Faults

By combining the failure traces expanded from stack traces and passing traces collected by running test cases, we obtain program spectra as illustrated in Figure 5.

|       | $f_1$ | ... | $f_i$ | $f_{i+1}$ | $f_{i+2}$ | $f_{i+3}$ | ... | $f_n$ | P/F |                    |
|-------|-------|-----|-------|-----------|-----------|-----------|-----|-------|------|-------------------|
| $T_1$ | 1     | ... | 1     | 0         | 0         | 1         | ... | 0     | fail | } failing traces  |
| ⋮     |       |     |       | ⋮         |           |           |     | ⋮     | ⋮    |                   |
| $T_m$ | 0     | ... | 0     | 1         | 0         | 0         | ... | 1     | fail |                   |
| $T_{m+1}$ | 1 | ... | 1     | 0         | 1         | 0         | ... | 1     | pass | } passing traces  |
| $T_{m+2}$ | 1 | ... | 1     | 0         | 1         | 0         | ... | 0     | pass |                   |
| ⋮     |       |     |       | ⋮         |           |           |     | ⋮     | ⋮    |                   |
| $T_t$ | 0     | ... | 1     | 0         | 0         | 1         | ... | 1     | pass |                   |

Figure 5. An example of program spectra.

In Figure 5, each line represents either a failing or passing trace and each column represents hit-record of a function in the trace.

The program spectra are then used to infer faulty entities by applying existing fault localization techniques. Examples of such techniques include *Ochiai*, *Tarantula* and *Jaccard*. We adapt *Ochiai* in our experiment, as early studies show that this technique can achieve slightly better fault localization performance [1]. However, our approach is not limited to *Ochiai*. We can use other techniques to locate faults leveraging expended failing and generated passing traces.

Our approach expands failing traces using static analysis. Expanded traces may include functions that will be never executed in the actual program executions. To improve fault localization performance, we propose the following two heuristics:

*Heuristics-1*: **Distance reweighting**

We have the following assumption for *Heuristics-1*: all functions in expanded failing execution traces are not equally important. Our empirical analysis finds that the functions that are closer to stack trace are more likely to contain faults. (We will describe the empirical analysis in Section VII.B.) Therefore, after conventional fault localization technique calculates a suspiciousness score to each function, we multiply the score with an extra coefficient $\beta$, which considers the distance from the stack trace.

$$\beta_i = \frac{n}{\sum_{j}^{n} dis_{i,j}}$$

, where $\beta_i$ denotes the distance weight for function $f_i$, $n$ is the number of stack traces whose expanded execution trace covers $f_i$, $dis_{i,j}$ represents the distance of $f_i$ to stack $s_j$, which is defined as the least number of calls from a function in the stack trace to $f_i$.

*Heuristics-2*: **Test coverage adjustment**

For software projects that have large code base, it is difficult to obtain 100% of the code test coverage rate. As we run test cases to collect the passing execution traces, the traces are likely to cover a limited percentage of code. There may be many program entities that are not covered by the passing execution traces. However, these entities may be included in the expanded failing execution traces since we statically expand the call stack. As a result, these entities will be regarded as highly suspicious ones by existing fault localization methods. This may lead to false positives. Consider the formula in *Ochiai* as an example, and suppose there is a program entity not covered by test cases but included in the failing execution traces.

$$S_i = \frac{a_{ef}}{\sqrt{(a_{ef} + a_{nf}) \cdot (a_{ef} + a_{ep})}}$$

$a_{ep}$ denotes the number of passing execution traces, thus for this program entity, $a_{ep}$ will be 0, which makes the denominator of the formula decline. As a result, its suspiciousness will increase.

To address this issue, we lower down the suspiciousness of program entities that are never covered by passing execution traces. Our heuristic revises the definition of $a_{ep}$ when the program entity is not covered by passing traces. When function $f_i$ is not covered by any passing execution trace, we define $a_{ep}(i)$ as the average value of all functions that appears in passing execution traces. The formula is listed as follows:

$$a_{ep}'(i) = \begin{cases} a_{ep}(i) & (a_{ep}(i) > 0) \\ \dfrac{\sum_i a_{ep}(i)}{\sum_i \delta(a_{ep}(i))} & (a_{ep}(i) = 0) \end{cases} \quad \delta(x) = \begin{cases} 1 & (x > 0) \\ 0 & (x \leq 0) \end{cases}$$

, where $a'_{ep}(i)$ denotes the modified number of appearances of $f_i$ in passing execution traces. We use $S'_i$ to denote the suspiciousness calculated by $a'_{ep}$. For example, for Ochai:

$$S'_i = \frac{a_{ef}}{\sqrt{(a_{ef} + a_{nf}) \cdot (a_{ef} + a'_{ep})}}$$

*Combining the Results:* For a function $f_i$, after obtaining its suspiciousness $S_i$ using existing fault localization methods, we adjust $S_i$ using the results of the heuristics as follows:

$$S''_i = S'_i \cdot \beta_i$$

, where $S''_i$ is the new suspiciousness score, $\beta_i$ denotes the distance weight for function $f_i$ calculated from *Heuristics-1*, while $S'_i$ is the suspiciousness using $a'_{ep}$ predicted by test coverage reweight coefficient.

## V. EXPERIMENTAL DESIGN

In this section, we describe our experimental design to evaluate the proposed approaches.

### A. Experimental Setup

We choose Mozilla Firefox[2], a large-scale open source web browser, as our subject. The project maintains publically available[3] crash data collected by a crash reporting system, Mozilla Crash Reporter.

We have collected crash reports for 23 releases of Firefox 3.6 and 14 releases of Firefox 4.0. Each Firefox release contains over 10k source files and 100k functions. We mine the actual crash fix location using the links between patch files in bug reports and crash reports. Since not all crash reports have links to bug reports, we only collected crash reports, for which 1) it has links to bug report and 2) the status of the bug report is either "RESOLVED FIXED" or "VERIFIED FIXED". With this condition, in total, we get 49,441 crash reports that correspond to 103 unique crashing faults. The statistics of our dataset is listed in Table II. For each bug report, we manually read its attached patch and identified actual fix locations.

TABLE II. THE FIREFOX DATASET

| Information | Firefox 3.6 | Firefox 4.0 |
|---|---|---|
| # Releases | 23 | 14 |
| # Source Code Files | 10K~11K | 11K~12K |
| # Lines of Code | 2224K~2351K | 2374K~2563K |
| # Faults | 61 | 42 |
| # Crash Reports | 27116 | 22325 |

---

[2] http://www.mozilla.org/firefox/
[3] http://crash-stats.mozilla.com

Firefox is a web browser, which takes web pages as test case for layout engine. We extracted more than 11K web pages from the Firefox project codebase and randomly selected 1,000 pages for testing. Scripts are written to invoke Firefox and executed the test cases automatically. We also used Mozmill[4], a capture and replay tool [24] containing 150 test scripts, for automatic UI interaction testing. In total, we used 1,150 test cases to generate the passing execution traces.

The experiment was conducted on a Dell PowerEdge 1950 server (4-core Xeon 5355 2.66 GHz processors, 8GB physical memory, and 400GB hard disk).

### B. Research Questions

To evaluate our approach, we design experiments to address the following research questions:

**RQ1:** How many crashing faults can be located using our approach?
**RQ2:** How much do heuristics improve fault localization performance?
**RQ3:** How do different statistical fault localization techniques perform in our framework?

RQ1 evaluates the effectiveness of our method that is described in Section IV. In Section IV.C, we have proposed two heuristics to improve the fault localization performance. RQ2 evaluates the effectiveness of the proposed heuristics. As stated in Section IV.B, our framework is independent of statistical fault localization techniques. RQ3 compares the effectiveness of these fault localization techniques when they are integrated into our framework.

### C. Evaluation Metrics

Fault localization approach produces a ranked list of all program entities with suspiciousness scores. Developers can examine the list and locate faulty entities. If a fault localization technique is effective, faulty entities should be given a relatively higher suspiciousness score and be ranked on top of the list.

We evaluate fault localization effectiveness by the percentage of faults that can be discovered by examining top $N (1 \leq N \leq 100)$ percent of functions. Thus a better fault localization technique allows users to examine less code while discover more faults. A figure for the cumulative number of faults vs. the percentage of code is also used to depict this metric [12] [31] [32].

In case a fault has multiple bug fixing points, we assume that after developers examining one bug fixing points, they can understand the problem and find out the rest of the bug fixing points for the fault via impact analysis.

### D. Dummy Guesser

An automated fault localization tool should perform at least better than random guessing. A dummy guesser simply output a random rank of all program entities that appear in the program spectra.

We model the performance of dummy guessing on Firefox dataset, and use it as a baseline.

---

[4] https://developer.mozilla.org/en/Mozmill_Tests

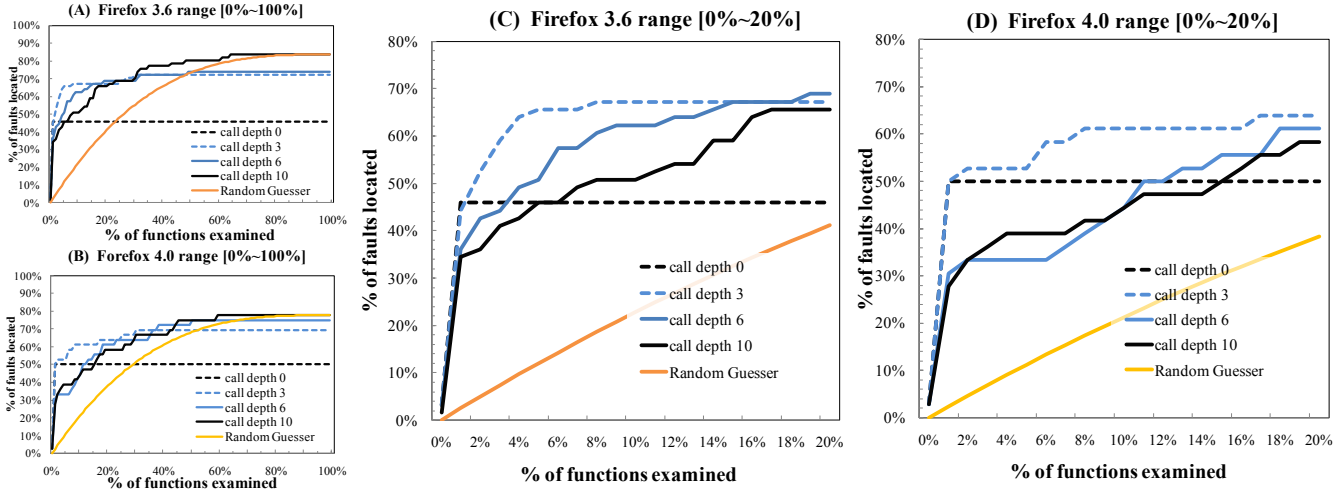

Figure 6. The evaluation results of the proposed approach.

In our experiment dataset (Table II), we have a total of 267 bug fixing points for 103 faults. Each fault has 2.59 bug fixing points on average. Suppose we have a program with *n* functions and one fault with three bug-fixing points for the fault. Following the assumption described in Section V.C, we assume developers can identify the rest of the bug fixing points after reviewing any one of the bug fixing points for a specific bug.

We then calculate the probability *Pr(m,n)* of locating a fault by examining *m* functions in the suspicious list generated by dummy guesser. *n* is the total number of functions.

$$\Pr(m,n) = \frac{A_{n-3}^{n-3} \cdot [(n-2) \cdot (n-1) \cdot (n) - (n-m-2) \cdot (n-m-1) \cdot (n-m)]}{A_n^n}$$

$$= 1 - \frac{(n-m-2) \cdot (n-m-1) \cdot (n-m)}{n \cdot (n-1) \cdot (n-2)}$$

where the denominator denotes the amount of all possible ranks. The numerator calculates the number of ranks that at least one faulty entity is in the top *m* recommended entities by extracting ranks that all three faulty entities are beyond the top *m*.

## VI. RESULTS

This section presents our experimental results by addressing the research questions.

*RQ1: How many crashing faults can be located using our approach?*

Figure 6 shows the performance of our approach, where x-axis is the percentage of functions examined and y-axis means the percentage of faults located.

For call depths 0~10, around 55K functions are included through the test case executions and the stack trace expansion.

The Random Guesser curves show the random guessing performance. The dashed line (call depth 0) represents the results using raw crash stack traces without expansion. For depth-0, 45% of faults can be located by examining less than 1% of functions in program spectra. However, no more faults can be covered when more code is examined. The reason is intuitive: depth-0 is synthesized from the original crash stack traces reported by Crash Reporting System and the passing execution traces collected by test case executions. Therefore only a small number of functions appear in failing execution traces. According to the formulas listed in Table I, only these functions will get non-zero suspiciousness. However, many faulty functions have been popped out from the stack trace before program crashes. As a result, only 45% of faults can be found even though developers examine all functions appeared in the program spectra.

Using deeper call depths, we can locate more faults. For example, when the depth is 3, examining 3% of functions can locate 52.46% faults; by examining 10% functions, developers can locate 67.21% faults, which is significant. Overall, the experiment results show that the proposed approach is promising: based only on crash stack traces, our approach can locate a large number of crashing faults by examining a relatively small percentage of functions.

Figure 6(A) also shows that after considering depth-10 function calls on Firefox 3.6, the proposed approach can cover 80% of all faults on both datasets. However, there is a tradeoff: higher recall leads to lower precision. When expanding 3 depths of function calls, programmers can locate 65.6% of faults by examining only 5% of functions. After considering 10 depths of function calls, programmers have to read 20% of functions to locate the same number of faults that can be found in depth-3.

*RQ2: How much do heuristics improve fault localization performance?*

Figure 7 shows the performance before and after applying the proposed heuristics when considering depths-10 function calls on Firefox 3.6 dataset. The conventional fault localization requires programmers to examine 10% of functions to locate only 4.76% of faults (as shown by the *original depth-10* curve in Figure 7(B)). After applying distance-reweighting heuristic, we observe a significant improvement over the conventional fault localizer. Programmers only need to examine less than 1% of functions to find out 35% of faults. By examining 10% of functions, 49.2% of faults can be located when both heuristics are applied, while 34.9% of faults are located when applying distance-reweighting only.

Overall, our heuristics significantly improves the fault localization performance.

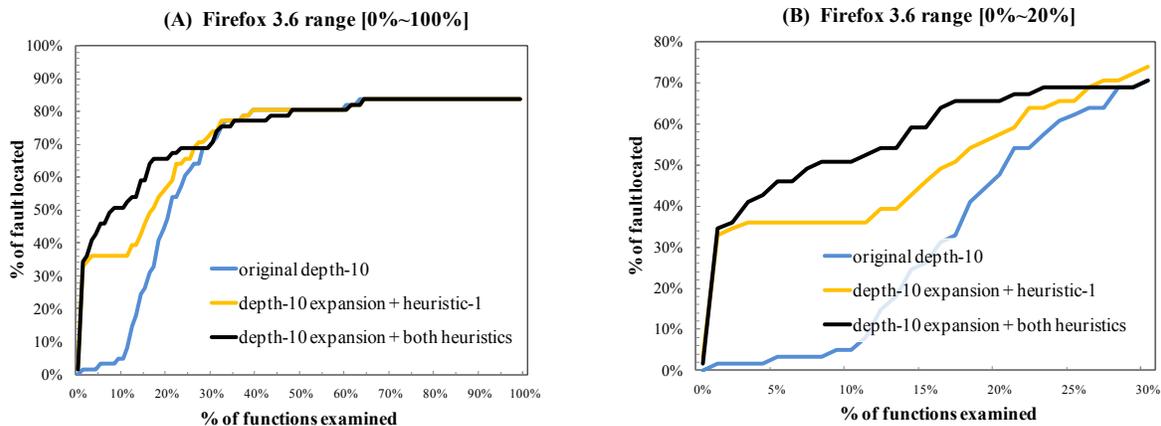

Figure 7. Effectiveness of using heuristics on all crashes.

*RQ3: How do different statistical fault localization techniques perform in our framework?*

Since our framework is not limited to a fault localization technique, this section compares the performance after applying three different statistical fault localization techniques with both heuristics in our framework. Table III shows the percentage of faults located by our method (with both heuristics) as the amount of code examined increase when depth-3 is considered. The result shows that these techniques can achieve similar performance using our framework.

We also conduct pair-wised t-tests for each pair of methods to see if some methods are statistically better than others. The result shows that there is no statistically significant difference between these techniques when they are integrated into our framework. These results indicate all three techniques yield similar performance using our framework.

TABLE III. COMPARING EXISTING METHODS IN OUR FRAMEWORK.

| % of functions examined | % of faults located | | |
|---|---|---|---|
| | *Tarantula* | *Jaccard* | *Ochiai* |
| 1% | 3.28% | 3.28% | 3.28% |
| 3% | 42.62% | 52.46% | 52.46% |
| 5% | 60.66% | 62.30% | 63.93% |
| 10% | 67.21% | 65.57% | 67.21% |
| 25% | 70.49% | 70.49% | 67.21% |
| 50% | 72.13% | 72.13% | 72.13% |
| 75% | 72.13% | 72.13% | 72.13% |
| 90% | 72.13% | 72.13% | 72.13% |

## VII. DISSCUSION

### A. Selecting and Learning the Optimal Depth

The choice of call depth affects the result of fault localization techniques. Generally, low depth can help locate faults by reading a relatively small percentage of code, but it limits the number of faults. Increasing the depth allows locating more faults, but it also includes more functions in suspiciousness list, making developer to investigate more codes. To measure this trade-off, we define the following metrics for evaluating the overall effectiveness of selecting different depths.

**Precision:** When faulty entities get higher ranks in the ranked list produced by fault localization, the precision is higher, therefore, we define precision as follows:

$$precision = \sum_i (1 - \frac{n_v(i)}{n_t}) \cdot w_d \qquad w_d = 1 - \frac{d}{d_{\max}}$$

For each faulty entity $i$ that is covered by the program spectra, $n_v(i)$ is the number of program entities viewed before locating fault $i$; $n_t$ is the number of program entities in the recommended list. The precision calculates the average percentage of code unnecessary to examine when locating faulty entities in the ranked list. In other words, when faulty entities are ranked top in the recommend list, the precision of will also be high. $w_d$ is a coefficient that considers the depth, since deeper depth means more functions added to the generated list and more functions to examine. $d$ is the current evaluating depth. $d_{max}$ is the maximum depth when execution traces stops expanding. Our empirical study shows that for Firefox project $d_{max}$ is 30(as illustrated in Figure 8 (A)).

**Recall:**
$$recall = \frac{n_{0\%\sim100\%}}{n_{faults}}$$

, where $n_{0\%\sim100\%}$ is the total number of faults covered by the program spectra, while $n_{faults}$ is the total number of faults. Therefore, Recall measures the percentage of faulty entities that are contained in the ranked list produced by fault localization.

**F-measure:**
$$F = \frac{2 \cdot precision \cdot recall}{precision + recall}$$

Figure 8(B) shows the Precision, Recall and F-measure values we computed for Firefox 3.6. It shows that when call depth increases, the F-measure goes up initially and then decreases. When call depth is three, the best overall performance is achieved.

For new projects, we can learn the optimal depth value from the historical projects. From historical projects, we can select the depth value that leads to the maximum F-measure as the optimal depth. We then use the obtained optimal call depth

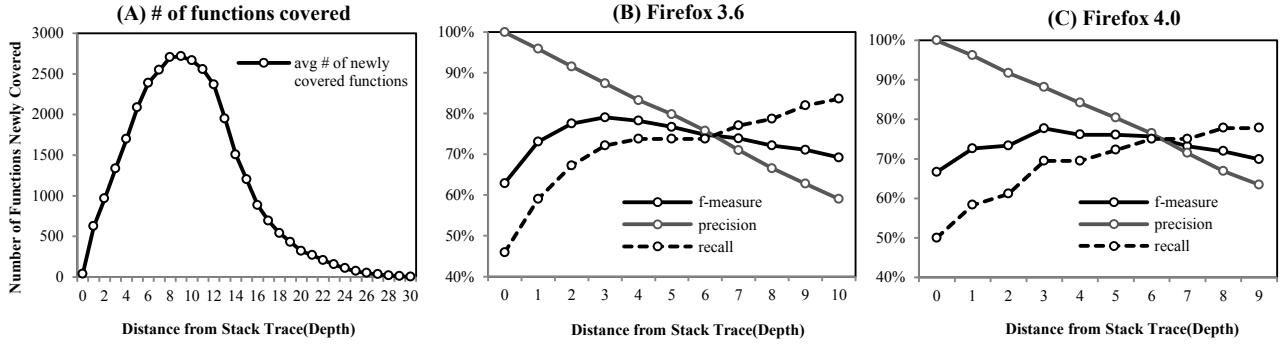

Figure 8. The selection of optimal depth.

to expand the call stack trace and perform fault localization for new projects.

To evaluate the method of learning the optimal depth, we use Firefox 3.6 crash data to build a fault localization model, and use the obtained optimal depth value learned from Firefox 3.6 to evaluate the fault localization performance on Firefox 4.0 cash data. The experimental results are shown in Figure 8. Based on the Firefox 3.6 dataset, the call depth 3 is selected as the optimal depth value (Figure 8(B)). As we can see in Figure 8(C), depth 3 still achieves the highest F-measure on Firefox 4.0 dataset.

*B. Where are the Faults?*

Our *Heuristic-1* is based on the assumption that functions that are closer to stack trace are more likely to contain faults. In this section, we empirically investigate this assumption. Table V shows the statistics about faults and newly expanded functions on each call depth. The number of newly found faults declines as the depth of function calls increases. For example, on Firefox 3.6 dataset, the original stack trace covers 32 faults; expanding original stack traces by one depth can cover 8 more faults. When one more depth is considered the number of newly found faults reduced to 5. Similar results can be observed on Firefox 4.0 dataset.

TABLE V. STATISTICS OF FUNCTIONS ON EACH CALL DEPTH

| | Depth | 0 | 1 | 2 | 3 | 4 | 5 | 6 | 7 | 8 | 9 |
|---|---|---|---|---|---|---|---|---|---|---|---|
| | # of newly expanded functions | 36 | 624 | 964 | 1332 | 1699 | 2086 | 2387 | 2547 | 2705 | 2714 |
| Firefox 3.6 | # of newly found faults | 32 | 8 | 5 | 3 | 1 | 0 | 0 | 2 | 1 | 2 |
| Firefox 3.6 | Cumulative # of faults found | 32 | 40 | 45 | 48 | 49 | 49 | 49 | 51 | 52 | 54 |
| Firefox 4.0 | # of newly found faults | 18 | 3 | 1 | 3 | 0 | 1 | 1 | 0 | 1 | 0 |
| Firefox 4.0 | Cumulative # of faults found | 18 | 21 | 22 | 25 | 25 | 26 | 27 | 27 | 28 | 28 |

Our findings are consistent with the results obtained by Adrian et al. [28], who did similar investigation on Eclipse[5] project. It shows that around 60% of bug reports contain bug fixing points in stack traces, and bug fixing points are more likely to be found in one of the top-10 stack frames. However, they only investigate the functions on the stack traces. Here we further study the bug fixing location by considering functions that are invoked from stack traces on various call depth. This empirical study supports the proposed *Heuristic-1* that more weights should be assigned to functions that are closer to the stack trace.

---

[5]http://www.eclipse.org/

*C. Data Quality Issues*

We presented a framework for locating crashing faults based on approximated program spectra, which is synthesized from expended failing traces and normal traces. Our program spectra may contain the following problems (Figure 9):

**Low test coverage rate**: In our approach, passing execution traces are generated from executing test suits. However, for large projects such as Firefox, it is often difficult to reach high test coverage.

**Noise on failing traces**: In our approach, a crash stack trace is expanded into approximate failing execution traces from a function call graph. Note that our function call graph is a static one that captures all possible call relationship between functions. Therefore, during stack trace expanding, many irrelevant functions are included into the execution traces. For example, if function A conditionally invokes either function B or function C, we include both B and C into the generated trace. The irrelevant functions could be considered as "noises" in the dataset if they are included into the execution traces.

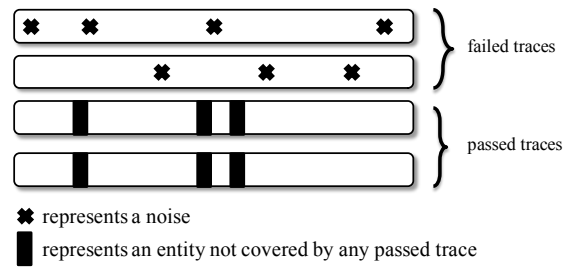

Figure 9. Imperfect Program Spectra.

To measure the impact of those negative facts in Figure 9, and evaluate the effectiveness of *Heurisitc-2* which copes with insufficient coverage issue on passing traces, we need a golden dataset for experiment. A golden dataset needs to have high-coverage-rate passing execution traces and noise-free failing traces. Due to the lack of those qualities on Firefox dataset, we use four subject programs provided by the Software-artifact Infrastructure Repository (SIR) [7]. These programs include flex, grep, gzip, and sed, which are all Unix utilities and real-life programs which have been widely adopted to evaluate many statistical fault localization techniques [11] [33] [1]. We investigate these four subject programs and find that 93.13% of statements in program spectra are covered by at least one passing trace. The failing traces are collected by statement-level instrumentation and execution and are noise-free. Table VI shows the descriptive statistics of each subject program, including the number of faults and the number of executable lines of code.

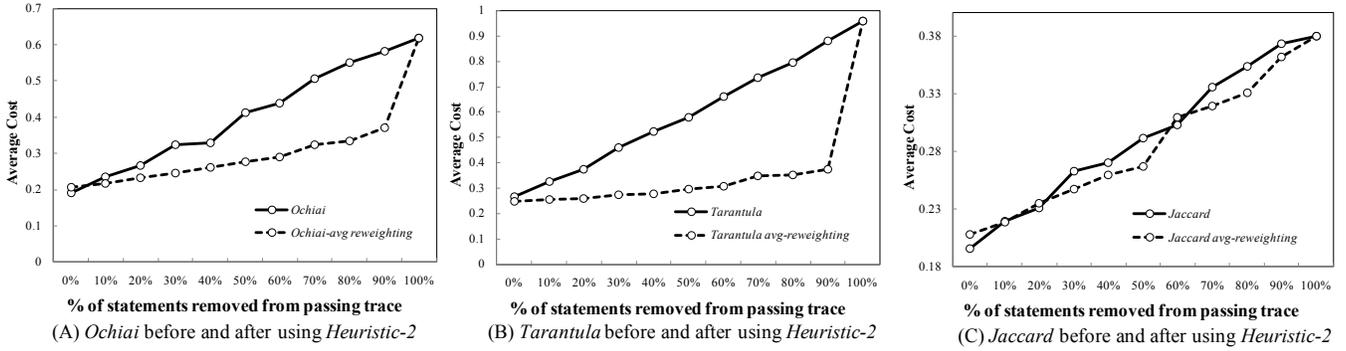

Figure 10. Effectiveness of *Heuristics-2* under different test coverage rates.

TABLE VI. SUBJECT PROGRAMS

| Program | LOC | Description | No. of Faults |
|---|---|---|---|
| flex | 8571-10124 | lexical parser | 21 |
| sed | 4756-9289 | text processor | 17 |
| grep | 8053-9089 | text processor | 17 |
| gzip | 4081-5159 | compressor | 55 |

Each subject program has different versions with different types of faults seeded. We activated these faults individually and executed the test suites to collect program spectra information. Following common practices on the subject programs [7], we excluded faults that cannot be revealed by any test case. In total, we used 110 single-fault versions to evaluate our technique.

*1) The Impact of Test Coverage Rate*

To observe the impact of test coverage rate on passing execution traces and measure the effectiveness of our proposed solution, we conduct a controlled experiment to simulate the scenario on the SIR benchmark suite [7].

To simulate certain test coverage rate, we randomly select a specified percent of program statements that appears in the clean passing traces, and change their hit-record from 1 to 0 as if they were never executed in passing execution traces. The more percent of statements removed from passing traces, the lower the test coverage (statement-level coverage) rate will be.

We evaluate the effectiveness of fault localization techniques using the average percentage of functions (i.e., average cost) that needed to be inspected in order to find a fault. Similar metric is also adopted in [27] [19] [32]. It is computed by the following formula:

$$p = \frac{r_b}{|S|} \qquad averageCost = \frac{\sum p}{n}$$

, where $r_b$ represents the rank of faulty function in generated inspecting list, $S$ represents the set of all functions that are executed by at least one test case and $n$ represents the number of faults. If average cost is low, it indicates that a small percentage of code needs to be inspected to find the faults.

Figure 10 shows that test coverage rates have a direct impact on the performance of conventional fault localization techniques. As described in Section IV.C, the reason is that, when we remove certain statements from passing execution trace, the suspiciousness of these statements would increase. As a result, the real faulty program entity's suspiciousness might be surpassed by those changed entities, and thus get a lower priority for examining.

We also compare the fault localization performance before and after applying *Heuristics-2* under different test coverage rates. As shown in Figure 10(A), when we remove 70% of statements from passing execution trace, original *Ochiai*'s average cost increases by 30%, while using the proposed heuristics the average cost raises 10%. *Tarantula* is also significantly improved by *Heurisitc-2* (Figure 10(B)) and *Jaccard* is slightly improved (Figure 10(C)). Here we observe a radical average cost increase of *Ochiai* and *Tarantula* with *Heurisitc-2* when we remove all statements (100%) from passed traces, since there is no candidate for reference which appears in the passing execution traces.

The results confirm that the proposed *Heuristic-2* significantly improves the fault localization performance under insufficient test coverage rates.

*2) Does the "Noises" Introduced by Stack Trace Expansion Affect the Fault Localization Performance?*

In this subsection, we describe the noise-resistant ability of fault localization methods that we found during the experiments. A small amount of noise does not affect the fault localization performance much.

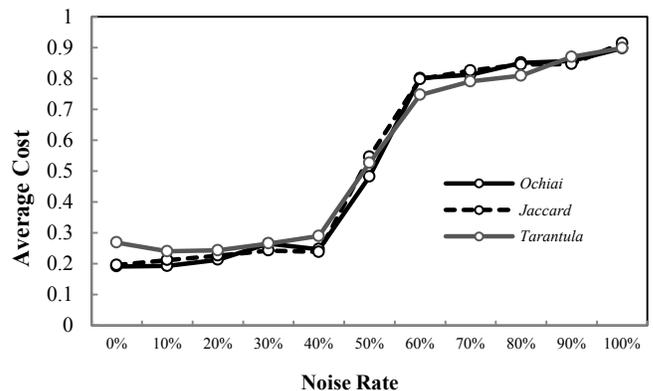

Figure 11. The average cost under various noise rate.

To explicitly demonstrate the noise-resistance of fault localization methods, we again use UNIX programs as "clean" dataset, since the actual failing traces of Firefox is absent. We create "noisy" dataset by flipping the execution status of randomly chosen program statements. For example, if a statement is not executed in one test case, we change its execution status in the trace from 0 to 1. We evaluate the performance of three commonly used fault localization methods (*Ochiai*, *Tarantula* and *Jaccard*) varying the rate of noise.

The results (Figure 11) show that even when 40% noise is inserted, the average cost does not change significantly. Only when the noise rate is over 50%, the performance of fault localization declines drastically. These results confirm that the fault localization methods have noise-resistance ability. Therefore, a small amount of noise introduced by our call stack expanding algorithm does not affect fault localization performance significantly.

## VIII. THREATS TO VALIDITY

We identify the following threats to validity:

- **Subject selection bias**: We use only Firefox project data in our experiments. Industrial source projects may have different crash properties and the same experiments on open source projects may yield different results.
- **Examining assumption**: In this paper, we measure the fault localization performance in terms of the first faulty function returned. If a fault affects multiple functions, we assume that after fixing the fault in the first relevant function, the developers are capable of identifying the root problem and locating the rest of functions via impact analysis.
- **Data collection**: We assume that one crash signature is caused by one fault. However, it is possible that there is more than one fault contributing to the same crash signature. In this case, stack traces not relevant to the fault could be included in the synthesized failing traces. This may affect fault localization performance.

## IX. RELATED WORK

In recent years, many studies have been dedicated to the analysis of crashes of real-world, large-scale software systems. These works include the construction of crash reporting system [9], the detection of duplicate crash reports [4] [14] and the prediction of crash-prone modules [8]. Our work also focuses on analyzing software crash data. Unlike the related work, we address the problem of locating faults when crashes happen.

Large software systems are often difficult to debug. Besides statistical fault localization techniques, many other techniques have also been proposed to facilitate debugging. For example, delta debugging [30] simplifies the failed test cases and yet preserves the failures, producing cause-effect chains and linking them to suspicious statements. Zhang et al. [31] applied program slicing techniques to fault localization by identifying a set of program entities that could affect the values of variables at a given program point. These techniques require either multiple re-executions of the program or extra program analysis efforts. Our approach applies statistical fault localization techniques and utilizes only available call stack trace information collected by crash report systems.

Liblit et al. [16] proposed a sparse sampling based statistical debugging method that can reduce the overhead of instrumentation in released program. Their sampling instrumentation technique incurs less than 5% slowdown at 1/1000 sampling rate. However, as they pointed out, lower sampling rate means that more sampling traces from users are required in order to observe the rare events (i.e., the observation of faulty entity execution). Therefore their method is more suitable for popular and widely-used software, while other approach only rely on crash stack traces collected by a crash reporting system. Furthermore, their approach requires users to execute specially instrumented software releases, while our approach requires only the normal releases of software.

Chilimbi et al. proposed an adaptive and iterative profiling method called Holmes [5] to locate post-release faults. Holmes also considers functions in stack trace that is closer to the crash point as more important. However, our method is different in that Holmes only considers functions in stack trace and reweights based on the distance to crash point, while the proposed *Heuristic-1* considers all functions in call graph and reweights based on the distance to stack trace. Also in Holmes it is used to select an initial function set for bootstrap of profiling iteration, while our method is used to reweight suspiciousness after fault localization.

Parnin et al. did an investigation [21] about the effectiveness of fault localization techniques by comparing programmers debugging time with and without automatic debugging tools; their results show that experts debug faster when using the generated list from fault localization. Participants also suggested that the primary benefit of such a tool is to point developers to the right direction. Our method combines crashing stack traces and passing execution traces to generate a ranking list which facilitates and guides debugging. We believe the proposed framework could help accelerating debugging crashes reported by end users.

## X. CONCLUSIONS

In this paper, we propose a novel framework for automatically locating crashing faults. Based on crash stack traces, our approach generates passing execution traces from test cases and failing execution traces from the stack trace extensions. We utilize existing fault localization techniques (such as *Ochiai*) as well as two heuristics to locate the crashing faults by contrasting the passing and failing execution traces.

The evaluations on Firefox 3.6 and 4.0 projects show that the proposed approach is effective to locate faulty functions. Using our framework, developers can locate 63.9% crashing faults by examining only 5% Firefox 3.6 functions in the spectra.

In future, we will explore if advanced static analysis (such as points-to analysis) can be applied to generate more precise execution traces. We will also evaluate our approach on more projects, including industrial projects.

The experimental tool and data used in this paper are available at:

*http://www.cse.ust.hk/~hunkim/neet/crash_fl.htm*